\documentclass[superscriptaddress,groupedaddress,nofootnoteinbib,12pt]{article} 
\pdfoutput=1
\usepackage{jcappub}
\usepackage{bm}
\usepackage{verbatim}

\usepackage{epsf}
\usepackage{graphicx,epsfig}
\usepackage{amsfonts}
\usepackage{amssymb}
\usepackage{xcolor}
\usepackage{textcomp}

\definecolor{mine}{rgb}{0.2,0.1,0.7}
\definecolor{bb}{rgb}{0.3, 0.5, 1}
\definecolor{bg}{rgb}{0.1, 0.1, 0.5}

\def\bkone{\mathbf k_1}
\def\bktwo{\mathbf k_2}
\def\bkthree{\mathbf k_3}
\def\bkfour{\mathbf k_4}

\newcommand{\sdelta}[1]{\!\delta (\mathbf{#1})}

\def\nn{\nonumber}

\newcommand\bk{\boldsymbol{k}}

\newcommand{\bea}{\begin{eqnarray}}
\newcommand{\eea}{\end{eqnarray}}
\newcommand\be{\begin{equation}}
\newcommand\ee{\end{equation}}
\newcommand\beq{\begin{equation}}
\newcommand\eeq{\end{equation}}

\def\ba{\begin{eqnarray}}
\def\ea{\end{eqnarray}}

\newcommand{\refeq}[1]{(\ref{#1})}

\def\M{M_{\rm Pl}}

\begin{document}

\title{The Trispectrum as a Diagnostic of Primordial Orthogonal non-Gaussianities}

\author[1,2]{S\'ebastien Renaux-Petel}
\affiliation[1]{Laboratoire de Physique Th\'eorique et Hautes Energies, University Paris 6, 4 place Jussieu, 75252 Paris, France.}
\affiliation[2]{Sorbonne Universit\'es, Institut Lagrange de Paris (ILP), 98 bis Bd. Arago, 75014 Paris, France.}
\emailAdd{srenaux@lpthe.jussieu.fr}
\vskip 4pt

\date{\today}


\abstract{In single-field inflationary models with a low sound speed, the orthogonal shape of the primordial bispectrum arises due to partial cancellations between equilateral-type shapes. This fact allows for a speed of sound $c_s$ as low as about 0.01, which is actually weakly preferred by WMAP data. For such values, the trispectrum, scaling like $1/c_s^4$, is of order $10^8$ and is therefore comparable to, and greater than, the 1$\sigma$ observational bound $t_{NL}^{\rm eq}=(-3.11 \pm 7.5) \times 10^6$. Hence, the trispectrum is already constraining inflationary mechanisms candidates for generating an orthogonal bispectrum at the level hinted in WMAP data. If this signal persists in imminent Planck data, most of the parameter space of the simplest effective field theory of inflation will be under observational pressure, while a dedicated analysis will be needed for the substantial fraction of parameter space where we show that a qualitatively new, orthogonal, trispectrum naturally arises.}


\maketitle


\section{Introduction}

The deviation from perfect Gaussian statistics of the primordial curvature perturbation $\zeta$ enables one to discriminate amongst the candidate physical mechanisms that produced the seed primordial fluctuations (see for instance \cite{Chen:2010xka,Byrnes:2010em,Wands:2010af,Barnaby:2010sq,Langlois:2011jt} for recent reviews). In this respect, it is fair to say that, despite significant efforts, the trispectrum has received considerably less attention than the bispectrum, in particular concerning data analysis. To our knowledge, three types of constraints on well motivated primordial trispectra are now available\footnote{Reference \cite{Fergusson:2010gn} constrains as well the non-primordial trispectrum signal induced by cosmic strings, as well as the primordial constant model, which provides a useful benchmark, but which has no physical motivation yet.}: constraints on $\tau_{NL}$ and $g_{NL}$ \cite{Smidt:2010ra,Fergusson:2010gn} -- which set the amplitude of the two different trispectra generated classically on super-Hubble scales in early-universe models with multiple degrees of freedom -- and constraints on $t_{NL}^{\rm eq}$ \cite{Fergusson:2010gn}, setting the amplitude of a representative `equilateral-type' trispectrum, generated by quantum interactions around Hubble crossing. The latter parameter is the equivalent for the trispectrum of $f_{NL}^{{\rm eq}}$ for the bispectrum, while the former are the counterparts of $f_{NL}^{{\rm loc}}$. Another important type of primordial bispectrum constrained by data, orthogonal non-Gaussianities \cite{Senatore:2009gt}, has up to now no counterpart at the level of the trispectrum.

The purpose of the present paper is two-fold. Our first aim is to show that an `orthogonal-type' trispectrum naturally arises in a significant fraction of parameter space in the simplest theoretical context (to be more accurate, this is a one-parameter family of trispectra as we will see). Our second, related, aim, is to point out the use of the trispectrum as a useful diagnostic for the appearance of a large bispectrum of the orthogonal type.

In particular, the final WMAP data \cite{Bennett:2012fp} indicated a 2.45$\sigma$ hint of orthogonal non-Gaussianities: $-445 < f_{NL}^{{\rm orth}}< -45\,\, (95 \% \, \,CL)$ (while showing no evidence of equilateral non-Gaussianities: $-221 < f_{NL}^{{\rm eq}}< 323\,\, (95 \%\,\, CL)$). When interpreted in terms of the effective field theory of inflation \cite{Cheung:2007st,Senatore:2009gt}, the orthogonal shape arises from partial  cancellations between otherwise equilateral shapes\footnote{Recently, a bispectrum with a significant overlap with the orthogonal shape was shown to arise in a different context \cite{Green:2013rd}.} (see \cite{RenauxPetel:2011dv,RenauxPetel:2011uk} for the first concrete realization of this mechanism and \cite{Renaux-Petel:2013ppa} for the calculation of the trispectrum in the same framework), leading to a smaller amplitude than the general estimate $f_{NL} \sim 1/c_s^2$, namely $f_{NL}^{{\rm orth}} ={\cal O} (\frac{0.01}{c_s^2})$. This fact allows for a speed of sound $c_s$ as low as about $0.01$, which is actually (weakly) preferred by current data \cite{Bennett:2012fp} (see Fig.~\ref{A-cs}). For such values of $c_s$, and unless inflation occurred in the region of parameter space where similar partial cancellations that leads to the orthogonal bispectrum arises at the level of the trispectrum, the amplitude of the trispectrum, scaling as $1/c_s^4$, is of order $10^8$ and hence is already comparable to (and greater than) current constraints $t_{NL}^{\rm eq}=(-3.11 \pm 7.5) \times 10^6$ obtained with WMAP data \cite{Fergusson:2010gn}. We will make this more quantitative in the body of this paper, but our message is simple: the trispectrum is already constraining inflationary mechanisms candidates for generating an orthogonal bispectrum at the level hinted in WMAP data. If this signal persists in imminent Planck data, most of the parameter space of the simplest effective field theory of inflation will be under strain, while a dedicated analysis of our orthogonal trispectrum signal will be needed for the remaining one.

The plan of our paper is as follows. In section \ref{sec:setup}, we use the effective field theory of inflation at the single-derivative level to parametrize the cubic and quartic action for fluctuations in low sound speed models. We then give the expression of the trispectrum generated in these models, using the results of Chen \textit{et al} \cite{Chen:2009bc} in the setup of k-inflation \cite{ArmendarizPicon:1999rj,Garriga:1999vw}, which is computationally equivalent. The overall amplitude of this trispectrum is fixed by the speed of sound $c_s$, while its shape depends on two-parameters: $A$, which determines the shape of the bispectrum, and $B$, which is unrelated in general. Section \ref{sec:shapes} is then dedicated to the study of the resulting 2-parameter family of shapes of trispectra in the region of parameter space $3.1 \lesssim A \lesssim 4.2$ where the orthogonal bispectrum is generated and where values of the speed of sound as low as ${\cal O}(0.01)$ are allowed. We define the region of parameter space where the trispectrum can be well represented by the `equilateral' one and is thus already constrained by present data, and represent the new shape that arises in the complementary region. We conclude in section \ref{sec:conclusion}. Eventually, the appendix \ref{Appendix:plots} collects a number of useful plots, while the appendix \ref{App:estimators} gives details about the construction of general estimators for the primordial trispectrum performed in Ref.~\cite{Regan:2010cn}.

\section{The trispectrum from low sound speed models}
\label{sec:setup}

In this section, we give the expression of the leading-order trispectrum generated in the simplest set-up of the effective field theory of inflation, which is computationally equivalent to k-inflation. Readers who are familiar with both the effective field theory of inflation and the trispectrum generated in low sound speed models can skip this section and proceed directly to section \ref{sec:shapes}.

\subsection{The effective field theory of inflation up to quartic order}

In this subsection, we briefly review the effective field theory of inflation developed in \cite{Cheung:2007st}, or to be more accurate the effective field theory of fluctuations generated in single-clock inflation. In such models with only one non-gravitational degree of freedom, it is always possible to choose a slicing such that surfaces of constant $t$ coincide with surfaces where the `clock' is unperturbed, \textit{i.e.} such that $\delta \phi(t, \boldsymbol{x})=0$ in the case of a scalar field clock. No explicit scalar fluctuations appear in this unitary gauge in which time diffeomorphisms have been fixed. The most generic effective action in this gauge can thus be built by allowing only metric operators invariant under the unbroken time-dependent spatial reparametrizations. It can then be shown that considering fluctuations around a spatially flat FLRW background amounts to studying the following action \cite{Cheung:2007st}
\bea
S=\int d^4x \sqrt{-g} \left[ \frac{\M^2}{2} R+\M^2 \dot H g^{00} -\M^2 (3 H^2+{\dot H}) +F(\delta g^{00},\delta K_{\mu \nu}, \delta R_{\mu \nu \rho \sigma};\nabla_{\mu};t) \right] \nn
\eea
where $H$ is the Hubble parameter, $\delta g^{00} \equiv g^{00}+1$, $\delta K_{\mu \nu}$ (respectively $\delta R_{\mu \nu \rho \sigma}$) is the fluctuation of the extrinsic curvature of constant time surfaces (respectively of the 4-dimensional Riemann tensor) and where $F$ starts quadratic in its arguments $\delta g^{00}$, $\delta K_{\mu \nu}$ and $\delta R_{\mu \nu \rho \sigma}$. The simplest effective field theory at lowest order in derivatives corresponds then to allowing operators involving powers of $\delta g^{00}$ only, namely, up to quartic order in fluctuations, 
\bea
F&=&\frac{1}{2} M_2(t)^4 (\delta g^{00})^2+\frac{1}{3!} M_3(t)^4 (\delta g^{00})^3+\frac{1}{4!} M_4(t)^4 (\delta g^{00})^4\,.
\eea
The true effectiveness of this approach relies on the gravitational analogue of the equivalence theorem for the longitudinal components of a massive gauge boson \cite{Cornwall:1974km}: the scalar degree of freedom can be explicitly reintroduced in the St\"uckelberg trick, thus restoring full time-diffeomorphism invariance, and it decouples from the gravitational sector at high enough energies, allowing to neglect the complications of the mixing with gravity. In this decoupling regime, the effect of the St\"uckelberg time diffeomorphism $t \to t+\pi(x)$ on $\delta g^{00}$ is simply
\be
\delta g^{00} \to -2 {\dot \pi}-{\dot \pi}^2+\frac{(\partial_i \pi)^2}{a^2}\,,
\ee
while one can neglect the terms introduced by the time dependence of the $M_n(t)$ at leading order in a slow-varying approximation, or equivalently by assuming that $\pi$ enjoys an approximate shift-symmetry. The resulting Lagrangian reads, up to quartic order in $\pi$:
\bea
S_{{\rm DL}}&=&\int d^4x \sqrt{-g} \left[ \M^2 \dot H (\partial_{\mu} \pi)^2+2 M_2^4\left({\dot \pi}^2-{\dot \pi} (\partial_{\mu} \pi)^2+\frac14((\partial_{\mu} \pi)^2)^2\right)
\right.
 \cr
&& \hspace{6.0em} +
  \left.
\frac{2 M_3^4}{3} \left(-2 {\dot \pi}^3+3 {\dot \pi}^2 (\partial_{\mu} \pi)^2  \right)+\frac{2M_4^4}{3}{\dot \pi}^4 \right]
\eea
where $(\partial_{\mu} \pi)^2 \equiv -{\dot \pi}^2+(\partial_i \pi)^2/a^2$ is evaluated on the background metric and $\pi$ is related to the curvature perturbation by the simple relation $\zeta=-H \pi$  at linear order and at leading order in a slow-varying approximation. A non-zero $M_2$ introduces both a reduced sound speed $c_s$, such that
\be
\frac{1}{c_s^2}-1 \equiv -\frac{2 M_2^4}{M_p^2 {\dot H}}\,,
\ee
and large cubic and quartic interactions. By introducing $A/c_s^2 \equiv -1+\frac23 \left(\frac{M_3}{M_2}\right)^4 $, following the WMAP notation in \cite{Bennett:2012fp}, and 
\be
\frac{B}{2c_s^4}\equiv 1-\left( \frac{M_3}{M_2} \right)^4+\frac12 \left( \frac{M_3}{M_2} \right)^8-\frac16 \left( \frac{M_4}{M_2} \right)^4 \,,
\ee
the decoupling Lagrangian can be cast in the form
\bea
S_{{\rm DL}}&=&\int d^4 x \sqrt{-g}\left[-\frac{M_P^2 \dot H}{c_s^2} \left({\dot \pi}^2-c_s^2\frac{(\partial_i \pi)^2}{a^2}\right)+\frac{M_P^2 \dot H}{c_s^2}\left(
\frac{{\dot \pi}(\partial_i \pi)^2}{a^2}+\frac{A}{c_s^2} {\dot \pi}^3 \right)
\right.
 \cr
&& \hspace{6.0em} 
  \left.
  -\frac{M_P^2 \dot H}{c_s^2}  \left( \frac{((\partial_i \pi)^2)^2}{4a^4} +\frac{3 A}{2 c_s^2} {\dot \pi}^2 \frac{(\partial_i \pi)^2}{a^2}+\frac{(9 A^2/4-B)}{c_s^4}  {\dot \pi^4}  \right)
 \right]
 \label{S-pi}
\eea
where we have kept leading-order terms in the interesting limit $c_s^2 \ll 1$ of a large bispectrum. As explained in \cite{Senatore:2009gt}, $A$ of order one is technically natural from the effective field point of view as the operators in ${\dot \pi}^3$ and $\dot \pi (\partial_i \pi)^2$ then introduce the same strong coupling scale. The same reasoning shows that $B$ of order one is technically natural as well. Note also that, as the effective field theory of inflation with operators involving only powers of $\delta g^{00}$ is computationally equivalent to k-inflation, it should not come as a surprise that the Lagrangian \refeq{S-pi} can be identified with the one in \cite{Chen:2006nt,Chen:2009bc} (see also \cite{Arroja:2009pd}), with the correspondence, at leading order in $1/c_s^2$:
\be
\frac{A}{c_s^2} \leftrightarrow - 2\frac{\lambda}{\Sigma}   \,, \qquad   \frac{B}{c_s^4} \leftrightarrow   \frac{\mu}{\Sigma}-9 \frac{\lambda^2}{\Sigma^2} \,.
\label{correspondence}
\ee
Note eventually that DBI inflation \cite{Silverstein:2003hf,Alishahiha:2004eh} simply corresponds to $A=-B=-1$ in our parametrization.

\subsection{The trispectrum}

Using the correspondence \refeq{correspondence}, the primordial trispectrum generated from the inflationary fluctuations Lagrangian \refeq{S-pi} can be simply read off from the equivalent result Eq.~(3.32) in \cite{Chen:2009bc}, namely
\bea
\langle \zeta(\bkone) \zeta(\bktwo) \zeta(\bkthree) \zeta(\bkfour) \rangle_c=(2 \pi)^9 {\cal P}_{\zeta}^3\,  \sdelta{\sum_i {\mathbf k_i}} \prod_{i=1}^4 \frac{1}{k_i^3} {\cal T}(k_1,k_2,k_3,k_4,k_{12},k_{14})
\label{def-T}
\eea
where $k_{ij}=   | \bk_{i}+\bk_{j}  |$,
\bea
{\cal T}(A,B)=\frac{1}{c_s^4} \left(\frac{A^2}{4}T_{s1}-\frac{A}{2} T_{s2}+T_{s3}-B T_{c1} \right)\,,
\label{T}
\eea
\be
T_{c1}=36 \frac{(k_1 k_2 k_3 k_4)^2}{(k_1+k_2+k_3+k_4)^5} 
\label{Tc1}
\ee
and where the explicit (lengthy) expressions of $T_{s1}, T_{s2}, T_{s3}$ can be found in \cite{Chen:2009bc}. Note that although the operators in  $((\partial_i \pi)^2)^2$ and ${\dot \pi}^2 (\partial_i \pi)^2$ in Eq.~\refeq{S-pi} are of the same order as the one in ${\dot \pi}^4$, they cancel in the quartic Hamiltonian at leading order in the small sound speed limit, leaving only the scalar exchange contributions $T_{si}$ and the contact-interaction trispectrum $T_{c1}$ generated from the operator in ${\dot \pi}^4$ (see \cite{Huang:2006eha} for the first calculation of $T_{c1}$).

\begin{figure}[!h]
  \center
  \includegraphics[width=0.7\textwidth]{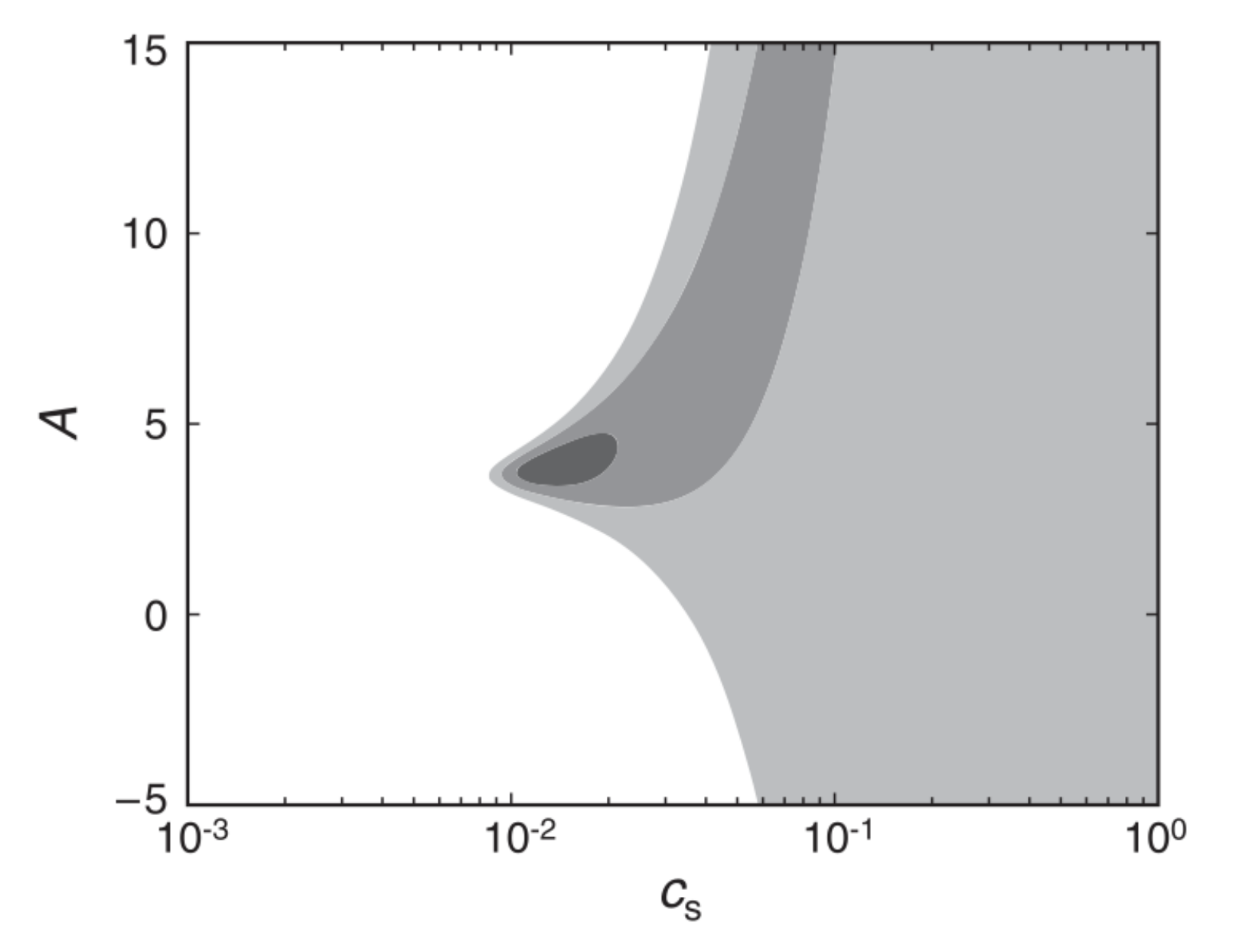}
      \caption{\label{A-cs} WMAP nine-year constraints: 1$\sigma$, 2$\sigma$ and $3\sigma$ confidence regions derived from the bispectrum on the sound speed $c_s$ and interaction coefficient $A$ in Eq.~\refeq{S-pi}. Figure taken from \cite{Bennett:2012fp}.}
\end{figure}

A detailed analysis of the shapes of the four constituent trispectra in \refeq{T} was performed in \cite{Chen:2009bc}, to which we refer the reader for more details (see also \cite{Arroja:2009pd}). Overall, they reached the conclusion that they all share very similar properties. In particular, as expected from trispectra of quantum origin generated around the time of Hubble crossing, they are the largest for the configurations where both the external and internal momenta are of similar magnitude, \textit{i.e.} near the regular tetrahedron (RT) limit where $k_1=k_2=k_3=k_4=k_{12}=k_{14}$. This similarity was used in \cite{Fergusson:2010gn} in which observational constraints on the simple and representative `equilateral-type' trispectrum $T_{c1}$ were derived (the only observational constraint on a trispectrum from quantum origin to date). However, we are interested here in the possible cancellations between these overall similar shapes, as we vary the parameters $(A,B)$, which could result in a trispectrum poorly correlated with $T_{c1}$, and to which current constraints would hence be blind. As we have explained in the introduction, of particular interest is the region (weakly) favored by WMAP nine-year data $3.1\lesssim A \lesssim 4.2$ where partial cancellations between the operators ${\dot \pi} (\partial_i \pi)^2$ and ${\dot \pi}^3$ in \refeq{S-pi} leads to a primordial bispectrum correlated with the orthogonal template at more than 80 \% \cite{Senatore:2009gt}, and in which a low sound speed of order $0.01$ is allowed \cite{Bennett:2012fp} (see Fig.~\ref{A-cs}). We present the results of such a study in the following section.

\section{`Equilateral' and orthogonal trispectra.}
\label{sec:shapes}

\subsection{Results}
\label{sec:results}

As we have just explained in the preceding section, we would like to assess where in parameter space the overall trispectrum \refeq{T} can/cannot be well represented by the simple `equilateral-type' trispectrum $T_{c1}$ constrained by data. Given the expression \refeq{T}, it is clear that, for any $A$, the trispectrum interpolates between highly correlated and highly anti-correlated with $T_{c1}$ as we vary $B$ from large negative to large positive values. A qualitatively new shape should hence arise in the neighborhood of  a particular value of $B$. The only questions are then: around which value? How narrow is this region? What the new shape looks like? And how much does the latter vary with $A$? in particular in the region  $3.1\lesssim A \lesssim 4.2$.

To answer these questions, one should ultimately resort to the scalar product defined by the estimator used to constrain the primordial trispectra,  for instance the correlator \cite{Hu:2001fa,Regan:2010cn}
\be
F[T,T']=\sum_{l_i,L} \frac{T^{l_1 l_2}_{l_3 l_4}(L) T^{' l_1 l_2}_{l_3 l_4}(L) }{(2L+1) C_{l_1} C_{l_2} C_{l_3} C_{l_4} }
\ee
between angular trispectra when using the CMB as the observational probe. Implementing such a correlator is however beyond the scope of this paper. We rather sticked to representing 2-dimensional slices of our trispectra in different representative limits, as was originally done in \cite{Chen:2009bc}. Note that we actually did make use of quantitative correlators, in Fourier-space, using reduced trispectra \cite{Regan:2010cn,Mizuno:2010by} or full trispectra \cite{Izumi:2011di}. However, we found these correlators to be somewhat misleading for our purpose, as they indicated widely different regions of parameter space as giving rise to an `orthogonal' trispectrum, in the sense of a trispectrum with a small correlation with $T_{c1}$. Moreover, when visually representing these candidate orthogonal trispectra, they appeared almost indistinguishable from $T_{c1}$. On the contrary, we believe that our procedure is trustworthy and sufficient to define where $T_{c1}$ represents well or not the overal trispectrum \refeq{T}, and, as we will explain below, we have quantitative arguments that support our findings.

To summarize the latter, we find that, for the three representative values $A=\lbrace 3.2,3.6,4 \rbrace$ in the interesting range $3.1\lesssim A \lesssim 4.2$, the total trispectrum \refeq{T} can be well represented by $T_{c1}$ only for
\bea
& \bullet &\,\,\   B\lesssim 5 \,\,{\rm and}\,\, B \gtrsim 11 \,\,(A=3.2) \label{range-Tc1-3.2} \\
 &\bullet& \,\,\   B\lesssim 5 \,\,{\rm and}\,\, B \gtrsim 14 \,\,(A=3.6)  \label{range-Tc1-3.6} \\
  &\bullet& \,\,\   B\lesssim 10 \,\,{\rm and}\,\, B \gtrsim 19 \,\,(A=4)  \label{range-Tc1-4}
\eea
\noindent while a qualitatively different shape arises in the complementary region, centered around the values
\bea
& \bullet &\,\,\   B \simeq 8.5 \,\,(A=3.2) \label{Orth-3.2} \\
& \bullet &\,\,\   B \simeq 11.5 \,\,(A=3.6) \\
& \bullet &\,\,\   B \simeq 14.5 \,\,(A=4) \label{Orth-4} 
\eea
\noindent Moreover, we find that the shapes of these various orthogonal trispectra depend very weakly on $A$ and can thus be well represented by (up to an overall multiplicative factor) 
\be
{\cal T}_{\rm orth}= 3.2\, T_{s1}-1.8\,T_{s2}+T_{s3}-11.5 \,T_{c1}\,,
\label{Torth}
\ee
corresponding to the values $A=3.6,B=11.5$. \\

Quantitatively, a simple measure of the amplitude of the trispectrum is given by the 
parameter $t_{NL}$ defined such that
\bea
\frac{1}{k^3} {\cal T}(k_1,k_2,k_3,k_4,k_{12},k_{14}) \xrightarrow[\rm limit]{\rm RT} t_{NL}\,,
\eea
where ${\rm RT}$ stands for the regular tetrahedron limit $k_1=k_2=k_3=k_4=k_{12}=k_{14}\equiv k$. Applied to \refeq{T}, this gives
\be
t_{NL}=\frac{1}{c_s^4} \left(0.062 A^2 - 0.210A +0.305 -0.035 B \right)\,.
\label{tNL}
\ee
The observational constraint relevant for our purpose in \cite{Fergusson:2010gn} is derived by assuming that ${\cal T}$ in Eq.~\refeq{def-T} can be well approximated by $T_{c1}$ in \refeq{Tc1}, \textit{i.e.} it puts a bound on the parameter $t_{NL}^{{\rm eq}}$ defined such that
 ${\cal T}=t_{NL}^{\rm eq}/ t_{NL}(T_{c1}) \times T_{c1}$. It is therefore applicable only in the ranges defined in Eqs.~\refeq{range-Tc1-3.2}-\refeq{range-Tc1-4} where the trispectrum becomes essentially proportional to $T_{c1}$. To make the link with the observational constraints on the bispectrum, let us apply this to the value of the speed of sound $c_s=0.013$\footnote{This value of $c_s$ is found using Eq.~57 in Ref.~\cite{Bennett:2012fp}.} such that $f_{NL}^{{\rm orth}}=-245$ (the central WMAP estimate) for the representative value $A=3.6$. One then finds
\be
t_{NL}=1.17 \times 10^7 \times (1-0.098 B)
\label{tNL-3.6}
\ee
which is comparable to (and actually greater than) the 1$\sigma$ constraint \cite{Fergusson:2010gn} 
\be
t_{NL}^{\rm eq}=(-3.11 \pm 7.5) \times 10^6\,.
\label{obs-bound}
\ee
Similar numbers are of course found in the range $3.1 \lesssim A \lesssim 4.2$: $t_{NL}=1.62 \times 10^7 \times (1-0.098 B)$ for $A=3.2$ and $t_{NL}=8.80 \times 10^6 \times (1-0.098 B)$ for $A=4$, under the same hypotheses. There is currently no point in performing a more detailed statistical analysis given the weak 2.45$\sigma$ hint of an orthogonal bispectrum, but we believe our message is clear: the trispectrum is already constraining candidate low sound speed inflationary models generating an orthogonal bispectrum at the level suggested in WMAP data. For instance, the result Eq.~\refeq{tNL-3.6} is $2.13\sigma$ (respectively $3.50\sigma$) away from the central value Eq.~\refeq{obs-bound} for $B=-1$ (respectively $B=-10$). Even a decrease in the error bar by a factor of a few, as expected from Planck \cite{Regan:2011zz}, could hence constrain these scenarios in a statistically significant way, provided of course that the orthogonal bispectrum signal is confirmed.

Note once again that Eq.~\refeq{tNL-3.6} can be meaningfully compared to the observational constraint \refeq{obs-bound} only for $B \lesssim 5$ and $B \gtrsim 14$. In the intermediate range, a shape qualitatively different from $T_{c1}$ arises to which current constraints are mostly blind. Relatedly, the fact that $t_{NL}$ in Eq.~\refeq{tNL-3.6} vanishes for $B=10.20$, so in the middle of this intermediate range, and close to the value $B=11.5$ at which we defined our representative orthogonal trispectrum, justifies \textit{a posteriori} our procedure and our findings. Indeed, $t_{NL}$ measures the amplitude of equilateral-type shapes, which peak around the regular tetrahedron limit. The fact that it vanishes does not indicate that no significant non-Gaussianities are generated, but rather that they can not be faithfully represented by the `equilateral' ansatz $T_{c1}$ in Eq.~\refeq{Tc1} The estimator $t_{NL}$ in Eq.~\refeq{tNL} vanishes as well at $B=7.76$ for $A=3.2$ and at $B=13.21$ for $A=4$, so again in the intermediate ranges that we defined and close to the values $B \simeq 8.5$ and $B \simeq 14.5$ at which defined the appearance of the orthogonal trispectrum.\\

Eventually, let us stress that the origin of our representative orthogonal trispectrum Eq.~\refeq{Torth} is not an ad-hoc orthogonalization procedure: by subtracting out the similarities between $T_{c1}$, $T_{s1}$, $T_{s2}$ and $T_{s3}$, one can indeed imagine the construction of a basis of the vector space spanned by these four trispectra constituted by $T_{c1}$ and three qualitatively different and mutually orthogonal trispectra. While mathematically correct, this procedure would be somewhat artificial. On the contrary, we have shown the natural appearance of trispectra qualitatively different from the equilateral one $T_{c1}$, and which can be represented by the template Eq.~\refeq{Torth}, in a substantial fraction of parameter space in the simplest theoretical context.

\subsection{Shapes of trispectra}

In this subsection, we show plots of $T_{c1}$ and of ${\cal T}(3.6,B)$ supporting the results stated above, namely that the latter can be represented to a good approximation by $T_{c1}$  for $B \lesssim 5$ and $B \gtrsim 14$, and that a trispectrum qualitatively different from $T_{c1}$ arises near $B=11.5$. We leave other plots, in particular for other values of $A$, to the appendix \ref{Appendix:plots}.

Our scale invariant trispectra are in general functions of $5$ variables, and in what follows, we follow Ref.~\cite{Chen:2009bc} and represent 2-dimensional slices of them in different representative limits. A slight difference is that we chose to plot the scale-independent quantities ${\cal {\tilde T}}={\cal T}/(k_1 k_2 k_3 k_4)^{3/4}$ rather than ${\cal T}$ itself (we have checked that similar conclusions are reached by using the two sets). The four limits we consider are:\\

  \textbullet \,\, \ The specialized planar limit, in which $k_1=k_3=k_{14}$, and the
    tetrahedron reduces to a planar quadrangle with \cite{Chen:2009bc} 
\begin{align}
      k_{12}=\left[
k_1^2+\frac{k_2 k_4}{2 k_1^2}\left( k_2 k_4 +
\sqrt{(4k_1^2-k_2^2)(4k_1^2-k_4^2)} \right) \right]^{1/2}~.
    \end{align}
   Shapes are then represented as functions of $k_2/k_1$ and $k_4/k_1$.

  \textbullet \,\, \  Near the double-squeezed limit: ${k}_3={k}_4=k_{12}$ and the tetrahedron
  is a planar quadrangle with \cite{Chen:2009bc} 
\begin{align}\label{planark2}
   k_2= \frac{\sqrt{k_1^2 \left(-k_{12}^2+k_3^2+k_4^2\right)- k_{s1}^2 k_{s2}^2+k_{12}^2 k_{14}^2+k_{12}^2 k_4^2+k_{14}^2
   k_4^2-k_{14}^2 k_3^2-k_4^4+k_3^2 k_4^2}}{\sqrt{2} k_4}~,
  \end{align}
where $k_{s1}$ and $k_{s2}$ are defined as
\begin{align}
&  k_{s1}^2\equiv 2\sqrt{(k_1 k_4+{\bf k}_1 \cdot {\bf k}_4)(k_1
k_4-{\bf k}_1 \cdot {\bf
  k}_4)}~,\nonumber\\ &
k_{s2}^2\equiv 2\sqrt{(k_3 k_4+{\bf k}_3 \cdot {\bf k}_4)(k_3
k_4-{\bf k}_3 \cdot {\bf
  k}_4)}~.
\end{align}
 Shapes are then represented as functions of $k_{4}/k_1$ and $k_{14}/k_1$.

  \textbullet \,\, \  The folded limit: $k_{12}=0$, hence $k_1=k_2$ and $k_3=k_4$. Shapes are then represented as functions of $k_4/k_1$ and $k_{14}/k_1$, and we assumed $k_4 < k_1$ without loss of generality.

 \textbullet \,\, \  The equilateral limit: $k_1=k_2=k_3=k_4$. Shapes are then represented as functions of $k_{12}/k_1$ and $k_{14}/k_1$.\\

\begin{figure}[!h]
  \center
  \includegraphics[width=1.0\textwidth]{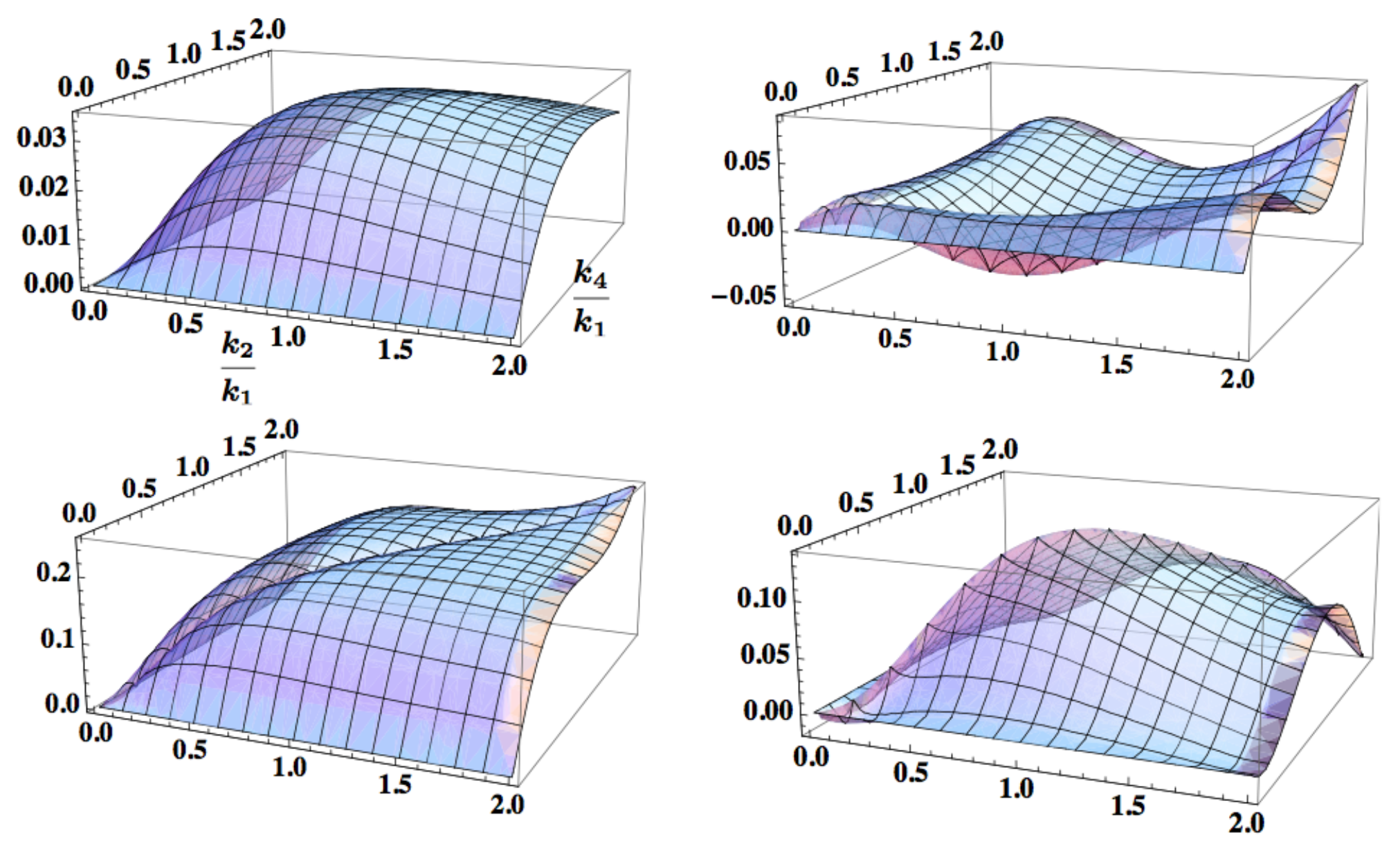}
      \caption{\label{SPL1}
         In this group of figures, we consider
      the specialized planar limit with $k_1=k_3=k_{14}$, and plot ${\tilde T}_{c1}$ (top left), ${\cal {\tilde T}}(3.6,11.5)$ (top right), ${\cal {\tilde T}}(3.6,5)$ (bottom left) and $-{\cal {\tilde T}}(3.6,14)$ (bottom right) as
      functions of $k_{2}/k_1$ and
    $k_{4}/k_1$.}
\end{figure}

\begin{figure}[!h]
  \center
  \includegraphics[width=1.0\textwidth]{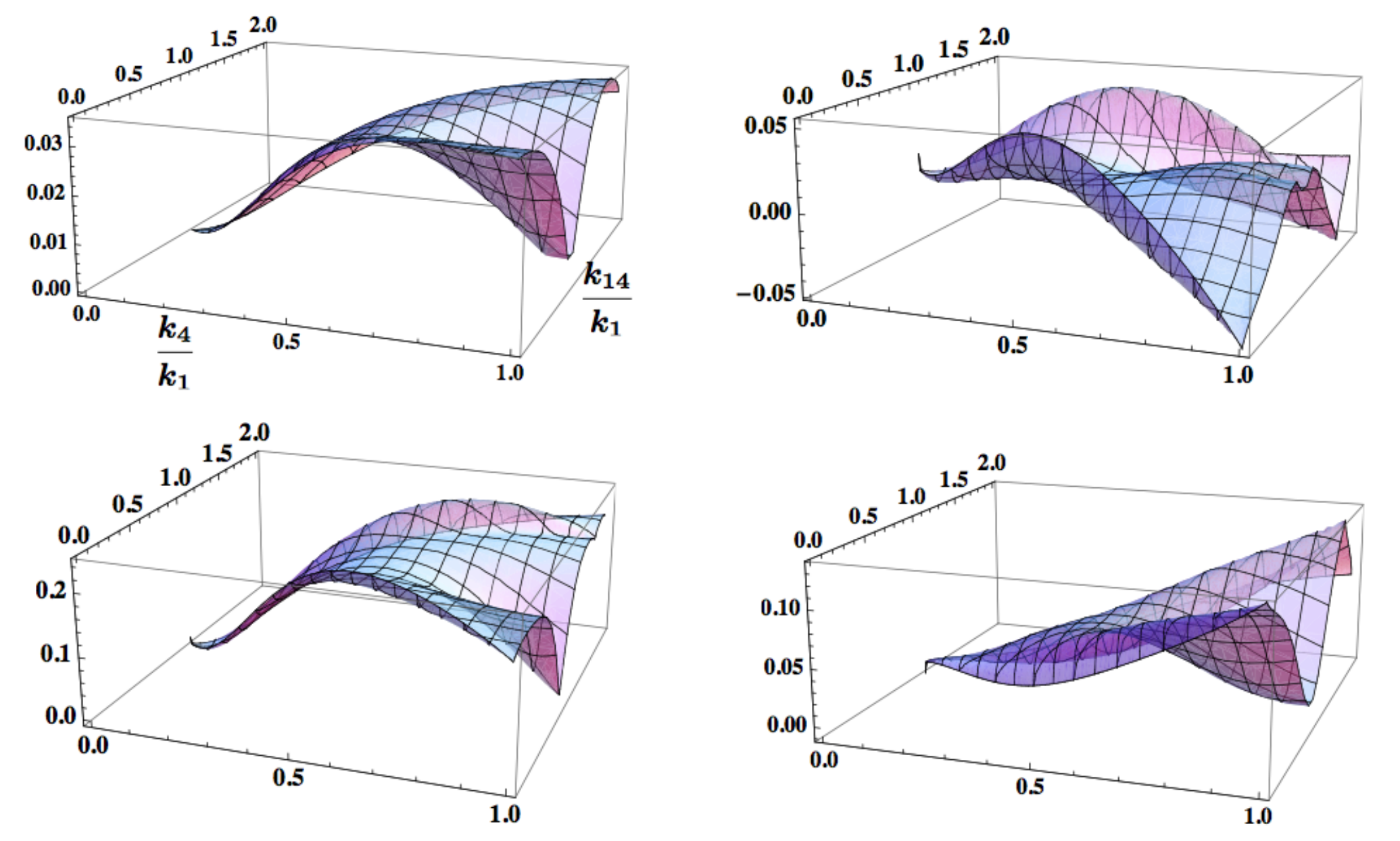}
      \caption{\label{DSL1} In this group of figures, we look
      at the shapes near
      the double squeezed limit: we consider the case where
      ${k}_3={k}_4=k_{12}$ and
      the tetrahedron
  is a planar quadrangle. We plot ${\tilde T}_{c1}$ (top left), ${\cal {\tilde T}}(3.6,11.5)$ (top right), ${\cal {\tilde T}}(3.6,5)$ (bottom left) and $-{\cal {\tilde T}}(3.6,14)$ (bottom right) as
      functions of $k_{4}/k_1$ and $k_{14}/k_1$.}
\end{figure}

\begin{figure}[!h]
  \center
  \includegraphics[width=1.0\textwidth]{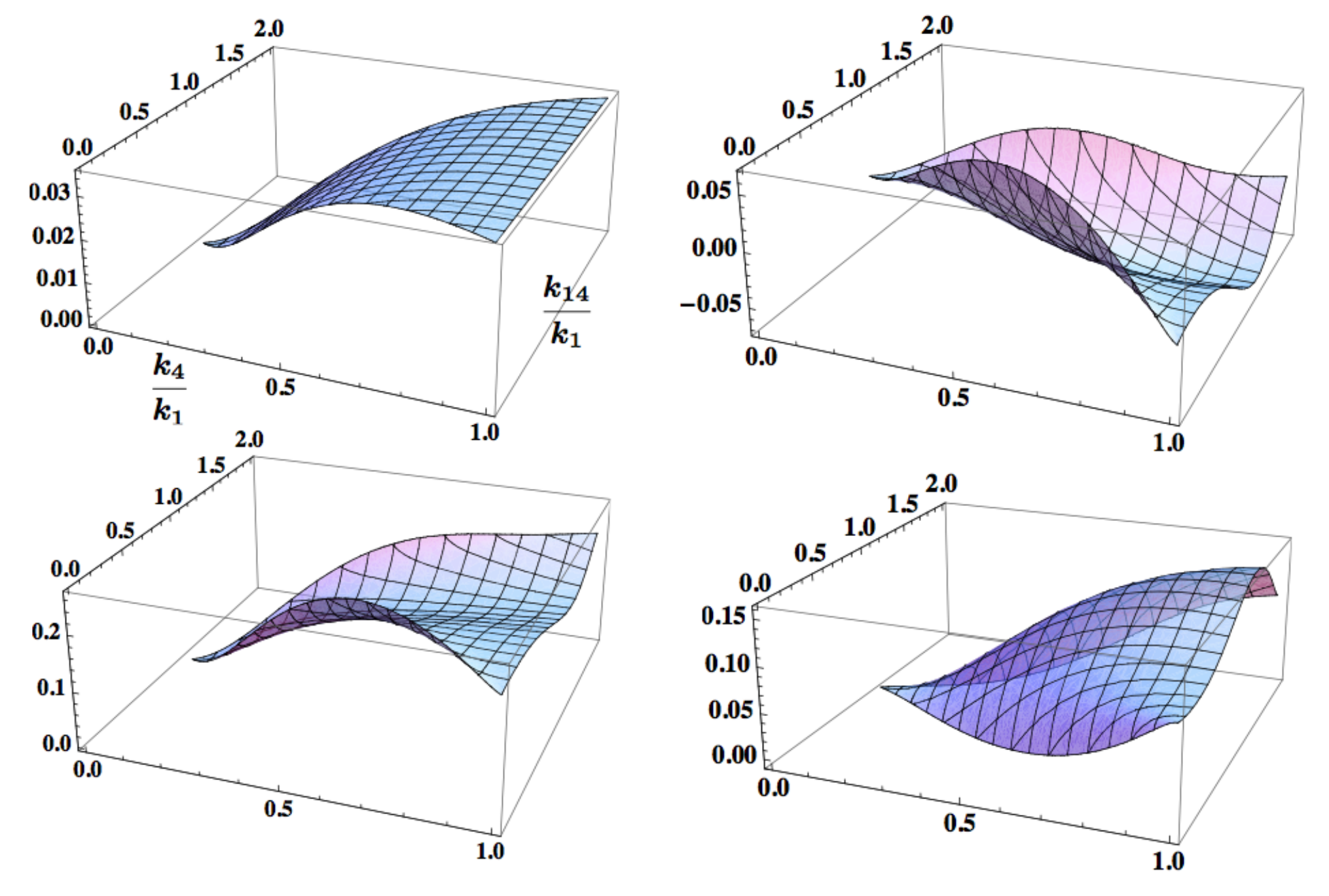}
      \caption{\label{Folded1} In this group of figures, we consider
      the folded limit
      $k_{12}=0$, and plot ${\tilde T}_{c1}$ (top left), ${\cal {\tilde T}}(3.6,11.5)$ (top right), ${\cal {\tilde T}}(3.6,5)$ (bottom left) and $-{\cal {\tilde T}}(3.6,14)$ (bottom right) as
      functions of $k_{4}/k_1$ and
      $k_{14}/k_1$.}
\end{figure}

\begin{figure}[!h]
  \center
  \includegraphics[width=1.0\textwidth]{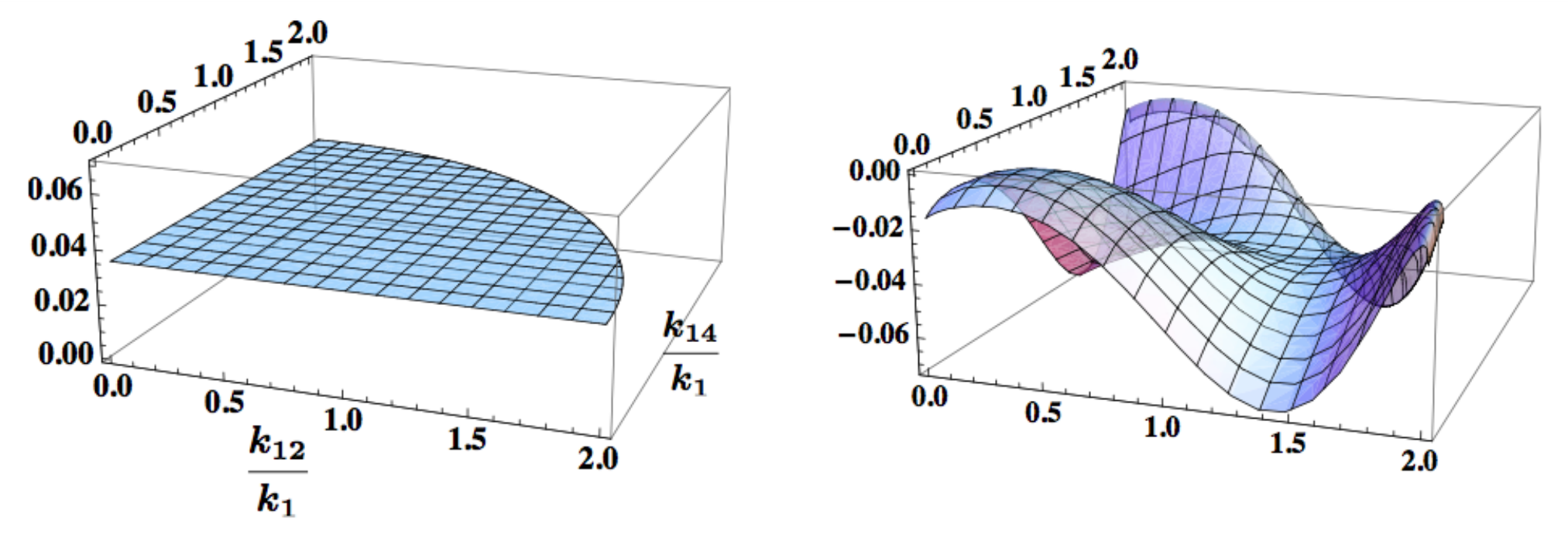}
      \caption{\label{Eq1} In this group of figures, we consider
      the equilateral limit $k_1=k_2=k_3=k_4$, and plot ${\tilde T}_{c1}$ (left) and ${\cal {\tilde T}}(3.6,11.5)$ (right) as
      functions of $k_{12}/k_1$ and
      $k_{14}/k_1$.}
\end{figure}

In Figs.~\ref{SPL1}, \ref{DSL1} and \ref{Folded1}, we represent the shape functions of ${\tilde T}_{c1}$ (top left), ${\cal {\tilde T}}(3.6,11.5)$ (top right), ${\cal {\tilde T}}(3.6,5)$ (bottom left)) and $-{\cal {\tilde T}}(3.6,14)$ (bottom right), in the specialized planar limit, near the double-squeezed limit and in the folded limit respectively. The shape functions are left white when the momenta do not form a tetrahedron. From these plots, it is evident that ${\cal {\tilde T}}(3.6,5)$ (respectively ${\cal {\tilde T}}(3.6,14)$) is well correlated (respectively anti-correlated) with ${\tilde T}_{c1}$, while ${\cal {\tilde T}}(3.6,11.5)$ is qualitatively different from it. Note that the scales are different in each plot, and that ${\cal {\tilde T}}(3.6,11.5)$ has both positive and negative values. In these three limits, the tetrahedron reduces to a planar quadrangle, which is the configuration probed by the CMB (see \cite{Regan:2010cn}). In Fig.~\ref{Eq1}, the non-planar, equilateral, limit, is also displayed. ${\tilde T}_{c1}$ is constant in this limit, so that its effect in Eq.~\refeq{T} is simply an overall shift in amplitude. For this reason, we represent only ${\tilde T}_{c1}$ (left) and ${\cal {\tilde T}}(3.6,11.5)$ (right).

Eventually, note that for $A \lesssim 3.1$ and $A \gtrsim 4.2$, the bispectrum is of equilateral type and observationally allowed values of the speed of sound $c_s$ are greater than in the range $3.1 \lesssim A \lesssim 4.2$, of order $c_s \gtrsim 0.1$ \cite{Bennett:2012fp} (see Fig.~\ref{A-cs}). The trispectrum \refeq{T} is then of maximum amplitude ${\cal O}(10^4)$ and, for such values, is not expected to be efficiently constrained by CMB data \cite{Creminelli:2006gc}. It is nonetheless interesting to investigate the appearance of orthogonal-type trispectra in this region of parameter space. We have not performed an exhaustive study of this, but the reasonings made in subsection \ref{sec:results} show that a trispectrum qualitatively different from $T_{c1}$ arises around a certain value of $B$ for any $A$, and that the cancellation of the estimator $t_{NL}$ in Eq.~\refeq{tNL} is a good estimate of where this arises, leading to the one-parameter family of approximate orthogonal trispectra
\bea
\frac{A^2}{4}T_{s1}-\frac{A}{2} T_{s2}+T_{s3}-8.66\left(1-0.69 A+0.20 A^2 \right) T_{c1} \,.
\eea
We have visually checked that this indeed the case, and moreover that outside the range $3.1 \lesssim A \lesssim 4.2$, these trispectra can differ substantially from our representative orthogonal trispectrum Eq.~\refeq{Torth}.

\section{Conclusions}
\label{sec:conclusion}

Amongst the three template bispectra constrained by the WMAP team, the orthogonal shape has been much less studied than local or equilateral non-Gaussianities. Although not statistically significant, the 2.45$\sigma$ hint of orthogonal bispectrum indicated by the final data \cite{Bennett:2012fp} hence deserves close attention. Observational consistency checks have been performed in \cite{Bennett:2012fp}, showing no obvious source of systematic contamination. The purpose of this paper was to show that the trispectrum is already providing another important consistency check, and that it actually puts under strain some of the simplest inflationary mechanisms generating such a signal. The reason is simple: when interpreted in the theoretical framework of the effective field theory of single-clock inflation, the WMAP signal favors a value of the sound speed of inflaton fluctuations of order ${\cal O}(0.01)$. For such low values, the trispectrum, scaling like $1/c_s^4$, is generically of order $10^8$, and hence is comparable to, and greater than, the current 1$\sigma$ observational bound $t_{NL}^{\rm eq}=(-3.11 \pm 7.5) \times 10^6$ \cite{Fergusson:2010gn}.

This limit is derived under the assumption that the trispectrum generated in low sound speed models can be well approximated by the simple `equilateral-type' shape generated by the operator ${\dot \pi}^4$. While this is true in a large fraction of parameter space, we have shown that there exists an non-negligible fraction of it in which qualitatively new shapes naturally arise, to which current constraints are almost blind. In the interesting region in which an orthogonal bispectrum is generated, an essentially unique orthogonal trispectrum Eq.~\refeq{Torth} can be generated, and it would be interesting to undertake an analysis of CMB data in search for such a signal.

\medskip

\begin{acknowledgments}

We would like to thank Xingang Chen, Paolo Creminelli, Keisuke Izumi, Kazuya Koyama, Shuntaro Mizuno, Guido W. Pettinari, Donough Regan, Filippo Vernizzi and Yi Wang for useful conversations related to the topic of this paper, and especially Keisuke Izumi, Shuntaro Mizuno and Donough Regan for useful explanations about their primordial trispectrum correlators, and Guido W. Pettinari for useful comments on a draft version of this paper. This work was supported by French state funds managed by the ANR within the Investissements d'Avenir programme under reference ANR-11-IDEX-0004-02.

\end{acknowledgments}

\appendix

\section{Comparison between orthogonal trispectra}
\label{Appendix:plots}

\begin{figure}[!h]
  \center
  \includegraphics[width=1.0\textwidth]{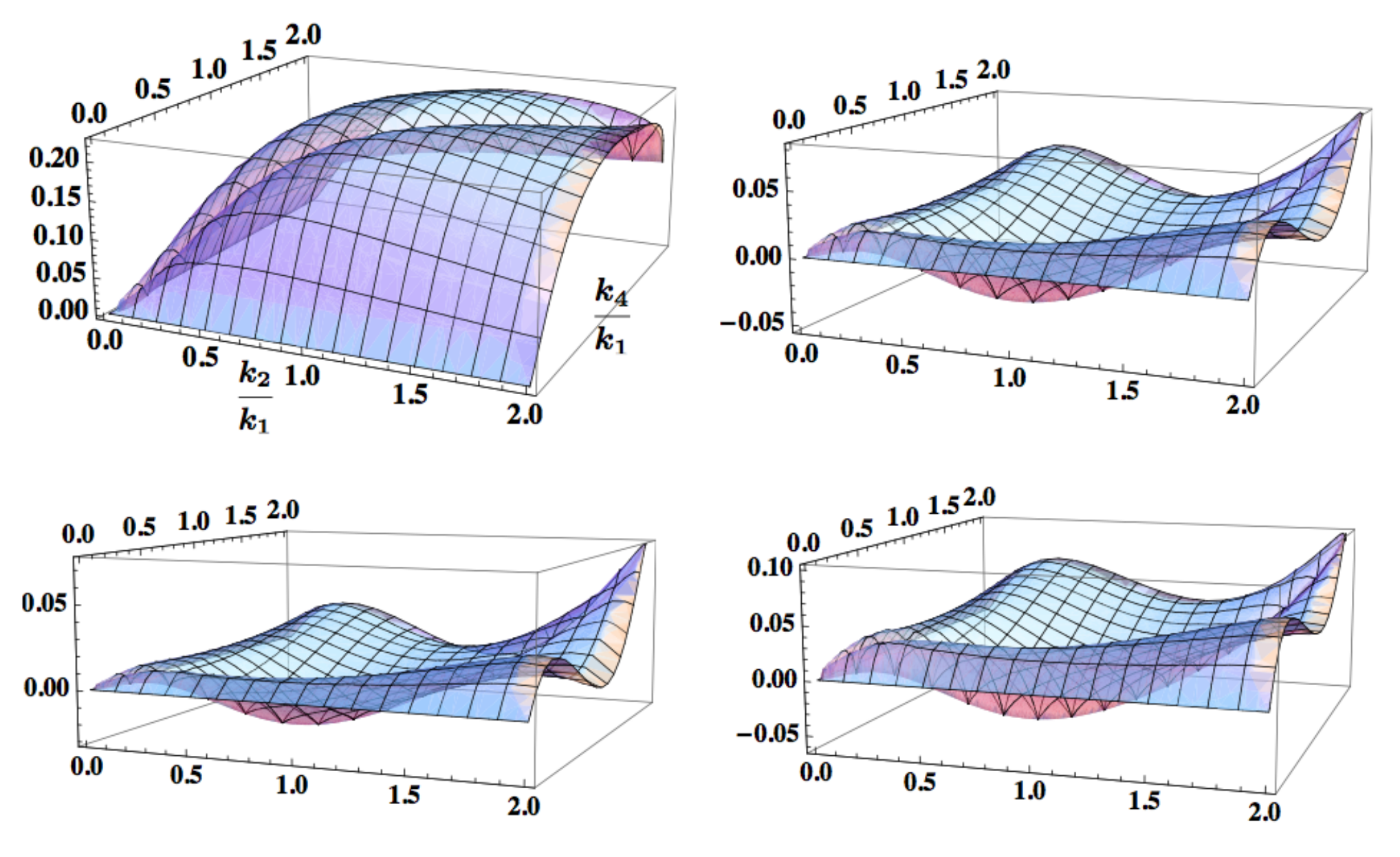}
      \caption{\label{SPL2}   In this group of figures, we consider
      the specialized planar limit with $k_1=k_3=k_{14}$, and plot ${\tilde T}_{s1}$ (top left), ${\cal {\tilde T}}(3.6,11.5)$ (top right), ${\cal {\tilde T}}(3.2,8.5)$ (bottom left) and ${\cal {\tilde T}}(4,14.5)$ (bottom right) as
      functions of $k_{2}/k_1$ and
    $k_{4}/k_1$.}
\end{figure}

\begin{figure}[!h]
  \center
  \includegraphics[width=1.0\textwidth]{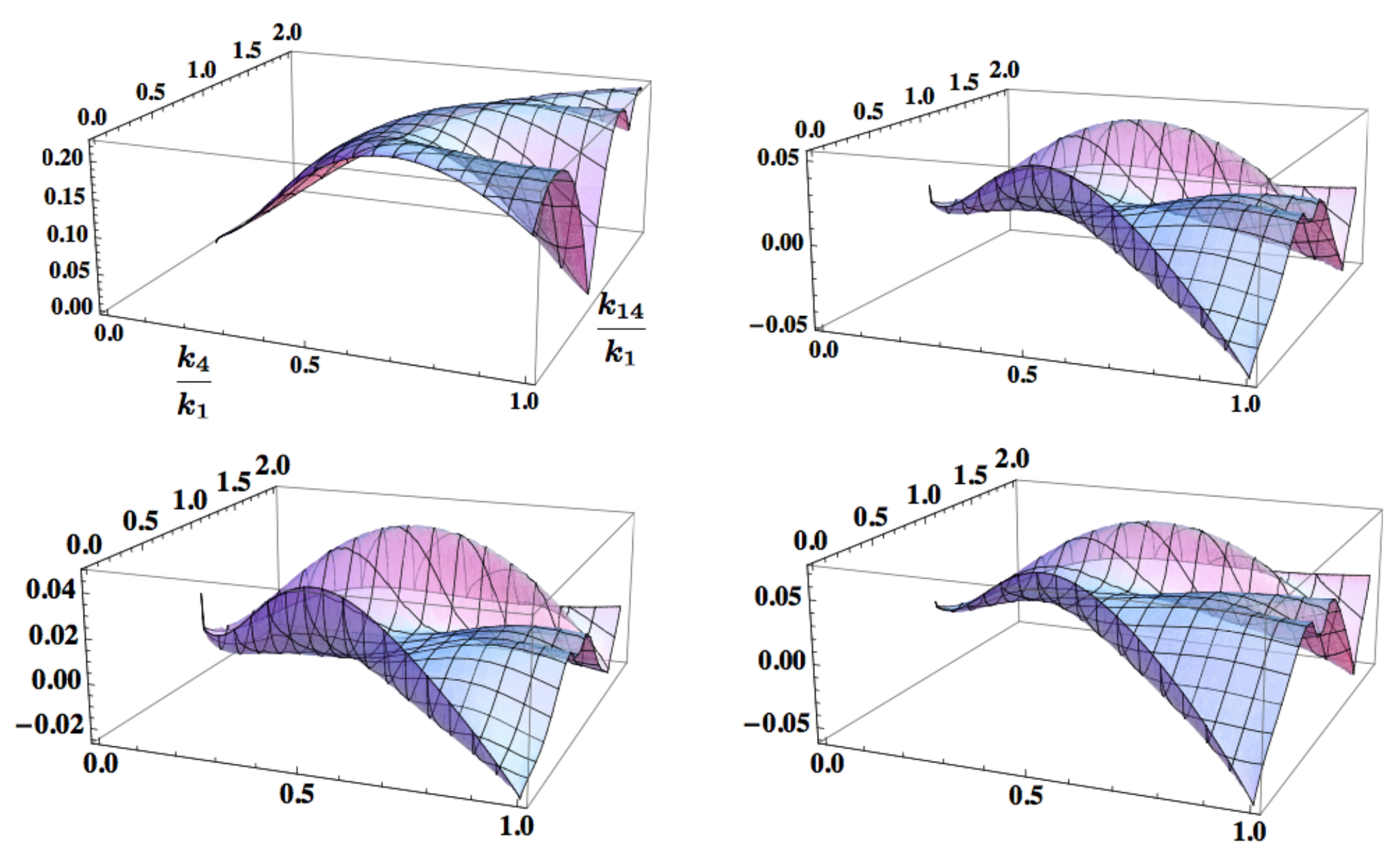}
      \caption{\label{DSL2} In this group of figures, we look
      at the shapes near
      the double squeezed limit: we consider the case where
      ${k}_3={k}_4=k_{12}$ and
      the tetrahedron
  is a planar quadrangle. We plot ${\tilde T}_{s1}$ (top left), ${\cal {\tilde T}}(3.6,11.5)$ (top right), ${\cal {\tilde T}}(3.2,8.5)$ (bottom left) and ${\cal {\tilde T}}(4,14.5)$ (bottom right) as
      functions of $k_{4}/k_1$ and $k_{14}/k_1$.}
\end{figure}

\begin{figure}[!h]
  \center
  \includegraphics[width=1.0\textwidth]{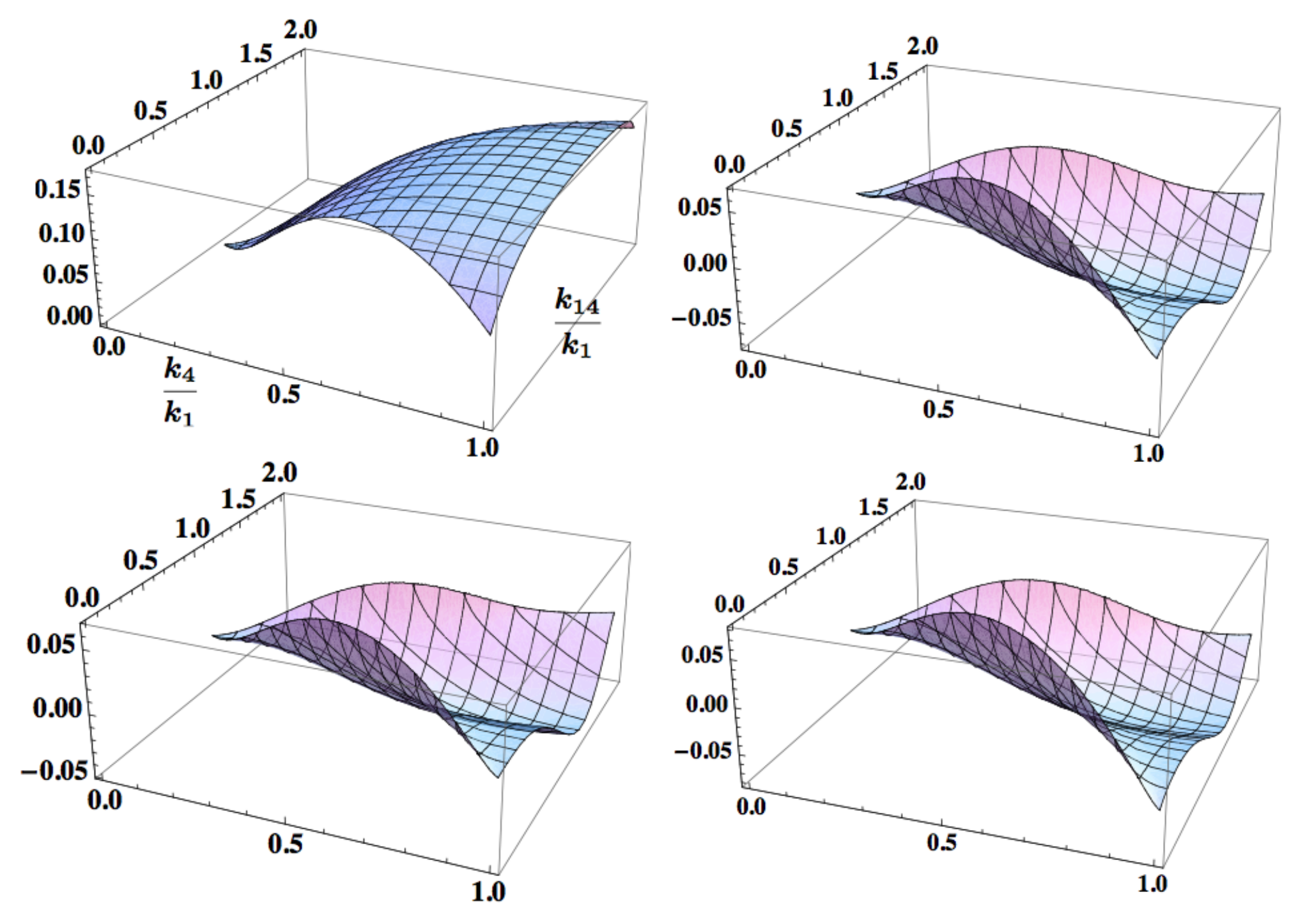}
      \caption{\label{Folded2} In this group of figures, we consider
      the folded limit
      $k_{12}=0$, and plot ${\tilde T}_{s1}$ (top left), ${\cal {\tilde T}}(3.6,11.5)$ (top right), ${\cal {\tilde T}}(3.2,8.5)$ (bottom left) and ${\cal {\tilde T}}(4,14.5)$ (bottom right) as
      functions of $k_{4}/k_1$ and
      $k_{14}/k_1$.}
\end{figure}

\begin{figure}[!h]
  \center
  \includegraphics[width=1.0\textwidth]{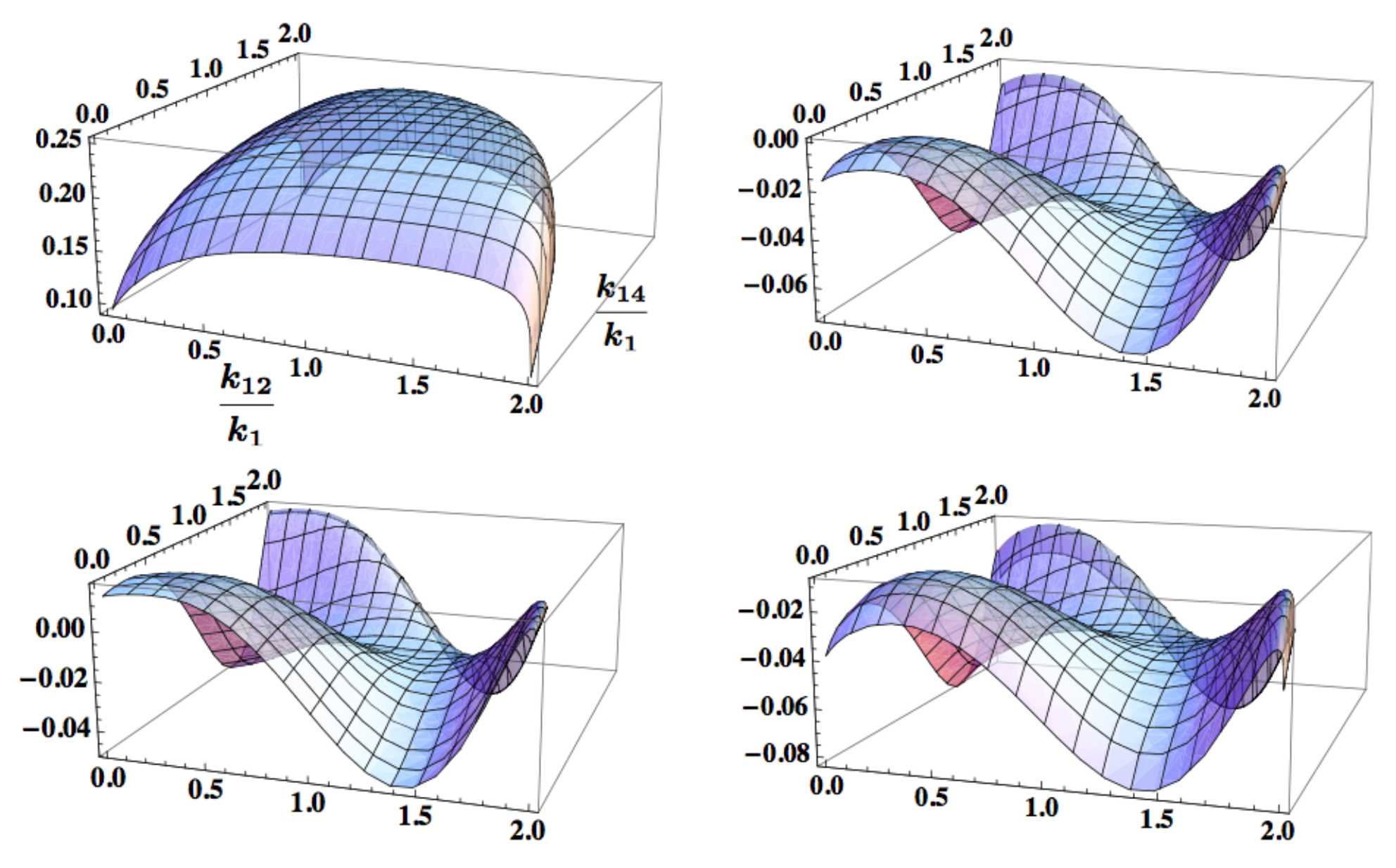}
      \caption{\label{Eq2} In this group of figures, we consider
      the equilateral limit $k_1=k_2=k_3=k_4$, and plot ${\tilde T}_{s1}$ (top left), ${\cal {\tilde T}}(3.6,11.5)$ (top right), ${\cal {\tilde T}}(3.2,8.5)$ (bottom left) and ${\cal {\tilde T}}(4,14.5)$ (bottom right) as
      functions of $k_{12}/k_1$ and
      $k_{14}/k_1$.}
\end{figure}

We have explained in subsection \ref{sec:results} that the trispectrum Eq.~\refeq{Torth} represents well the orthogonal trispectra Eqs.~\refeq{Orth-3.2}-\refeq{Orth-4} arising in the range $3.1 \lesssim A \lesssim 4.2$. In this appendix, we demonstrate this by showing, in Figs.~\ref{SPL2}, \ref{DSL2}, \ref{Folded2} and \ref{Eq2}, plots of ${\cal {\tilde T}}(3.2,8.5)$, ${\cal {\tilde T}}(3.6,11.5)$ and ${\cal {\tilde T}}(4,14.5)$ in the same limits as for Figs.~\ref{SPL1} to \ref{Eq1}. For comparison, we also represent ${\tilde T}_{s1}$, an `equilateral-type' shape different from ${\tilde T}_{c1}$, which we have represented in the other figures. The similarity between ${\cal {\tilde T}}(3.2,8.5)$, ${\cal {\tilde T}}(3.6,11.5)$ and ${\cal {\tilde T}}(4,14.5)$ is striking.

\section{Estimators}
\label{App:estimators}

For the sake of completeness, we present in this appendix an expression for the trispectrum estimator that can be used to constrain with CMB data the various non-Gaussian signals studied in this paper. We stress that no original material is presented here. In particular, for more details, we refer the reader to \cite{Regan:2010cn}, where the construction of this estimator has been carried out.

We use the same conventions and notations as in \cite{Regan:2010cn}, and consider the four-point function induced by a general non-Gaussian primordial gravitational potential $\Phi(\bk)$ in the temperature fluctuations of the CMB. The field $\Phi$ is simply related to the primordial curvature perturbation $\zeta$ by $\Phi=\frac{3}{5} \zeta$ and induces the multipoles $a_{lm}$ via a convolution with the transfer functions $\Delta_l(k)$ through the relation
\begin{eqnarray}
a_{lm}=4\pi (-i)^l \int \frac{d^3k}{(2\pi)^3} \Delta_l(k) \Phi(\mathbf{k}) Y_{lm}(\hat{\mathbf{k}}).
\end{eqnarray}

An important quantity for what follows is the primordial reduced trispectrum $\mathcal{T}_{\Phi}$, which is related to the primordial trispectrum by 
\begin{eqnarray}\label{Tconn}
\langle \Phi(\mathbf{k_1})\Phi(\mathbf{k_2})\Phi(\mathbf{k_3})\Phi(\mathbf{k_4}) \rangle_c=(2\pi)^3 \int d^3 K \delta (\mathbf{k_1+k_2+K})\delta (\mathbf{k_3+k_4-K})T_{\Phi}(\mathbf{k}_1,\mathbf{k}_2,\mathbf{k}_3,\mathbf{k}_4;\mathbf{K})\,, \nonumber
\end{eqnarray}
with
\begin{eqnarray}
T_{\Phi}(\mathbf{k}_1,\mathbf{k}_2,\mathbf{k}_3,\mathbf{k}_4;\mathbf{K})=&&P_{\Phi}(\mathbf{k}_1,\mathbf{k}_2,\mathbf{k}_3,\mathbf{k}_4;\mathbf{K})+\int d^3 K' [\delta (\mathbf{k_3-k_2-K+K'})P_{\Phi}(\mathbf{k}_1,\mathbf{k}_3,\mathbf{k}_2,\mathbf{k}_4;\mathbf{K'})\nonumber \\
&&+\delta (\mathbf{k_4-k_2-K+K'})P_{\Phi}(\mathbf{k}_1,\mathbf{k}_4,\mathbf{k}_3,\mathbf{k}_2;\mathbf{K'}) ]
\end{eqnarray}
and 
\begin{eqnarray}
P_{\Phi}(\mathbf{k}_1,\mathbf{k}_2,\mathbf{k}_3,\mathbf{k}_4;\mathbf{K})&=&\mathcal{T}_{\Phi}(\mathbf{k}_1,\mathbf{k}_2,\mathbf{k}_3,\mathbf{k}_4;\mathbf{K})+\mathcal{T}_{\Phi}(\mathbf{k}_2,\mathbf{k}_1,\mathbf{k}_3,\mathbf{k}_4;\mathbf{K})\nonumber\\
&+&\mathcal{T}_{\Phi}(\mathbf{k}_1,\mathbf{k}_2,\mathbf{k}_4,\mathbf{k}_3;\mathbf{K})+\mathcal{T}_{\Phi}(\mathbf{k}_2,\mathbf{k}_1,\mathbf{k}_4,\mathbf{k}_3;\mathbf{K}).
\end{eqnarray}
The rationale behind these definitions lies in the fact that we need only consider the reduced trispectrum $\mathcal{T}_{\Phi}$ from one particular arrangement of the various wavevectors and form the other contributions by permuting the symbols.
\begin{figure}[htp]
\centering 
\includegraphics[width=102mm]{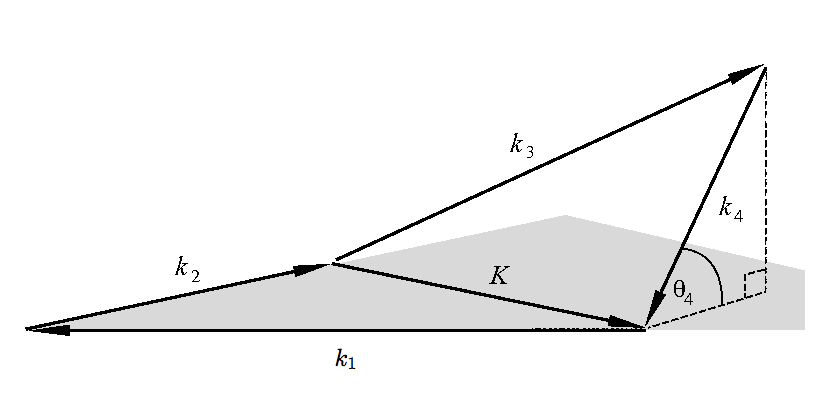}
\caption{Quadrilateral defined by the four wavenumbers ${\boldsymbol k}_i$, the diagonal $K$, and the 
angle $\theta_4$ out of the plane of the first triangle. Figure taken from \cite{Regan:2010cn}.}
\label{fig:Tquad2}
\end{figure}
Furthermore, it turns out to be useful to expand $\mathcal{T}_{\Phi}$ as a Legendre series in terms of the angle $\theta_4$, which represents the deviation of the quadrilateral defined by the ${\boldsymbol k}_i$ from planarity (see Fig.~\ref{fig:Tquad2}):
\begin{eqnarray}\label{expansion}
\mathcal{T}_{\Phi}(k_1,k_2, k_3,k_4; K, \theta_4)=\sum_{n=0}^{\infty}\mathcal{T}_{\Phi,n}(k_1,k_2, k_3,k_4; K) P_n(\cos\theta_4)\,,
\end{eqnarray}
where $K=| \mathbf{K} |$. It can indeed be shown that the CMB only probes and constraints the planar $(n=0)$ component of the trispectrum, \textit{i.e.} $\mathcal{T}_{\Phi,0}(k_1,k_2, k_3,k_4; K)$ \cite{Regan:2010cn}.\\

In the idealised limit where sky cuts and inhomogeneous noise can be neglected (see \cite{Regan:2010cn} for the more general case), the expression for the trispectrum estimator is
\begin{eqnarray}\label{Estimator}
\mathcal{E}=\frac{1}{N_T}\sum_{l_i m_i}\frac{\langle a_{l_1m_1}a_{l_2m_2}a_{l_3m_3}a_{l_4m_4}\rangle_c \left(a_{l_1 m_1}^{\rm{obs}}a_{l_2 m_2}^{\rm{obs}}a_{l_3 m_3}^{\rm{obs}}a_{l_4 m_4}^{\rm{obs}}\right)_c}{C_{l_1}C_{l_2}C_{l_3}C_{l_4}}
\end{eqnarray}
where the normalisation factor $N_T$ is given by 
\begin{eqnarray}\label{Normal}
N_T=\sum_{l_i, L}\frac{(T^{l_1 l_2}_{l_3 l_4}(L))^2}{(2L+1)C_{l_1}C_{l_2}C_{l_3}C_{l_4}}
\end{eqnarray}
in terms of the angular trispectrum $T^{l_1 l_2}_{l_3 l_4}(L)$. Rendering the estimation of $\mathcal{E}$ computationally tractable requires, similarly to the case of the bispectrum \cite{Fergusson:2006pr,Fergusson:2008ra,Fergusson:2009nv}, the introduction of a separable representation of the planar component of the (rescaled) reduced trispectrum $\mathcal{T}_{\Phi,0}$:
\begin{eqnarray}
(k_1 k_2 k_3 k_4)^2 K \mathcal{T}_{\Phi,0}=\sum_{m}  \alpha_{m}^{\mathcal{Q}}\mathcal{Q}_m (t,s,x,y,z)\,,
\end{eqnarray}
where
\begin{eqnarray}
\mathcal{Q}_m(t,s,x,y,z)= q_p(s)q_r(x)q_s(y)q_u(z) r_v(t)
\end{eqnarray}
is formed by products of one dimensional-functions of  $s=k_1/k_{\rm{max}},x=k_2/k_{\rm{max}},y=k_3/k_{\rm{max}},z=k_4/k_{\rm{max}}, t=K/k_{\rm{max}}$ and $m=\{p,r,s,u,v\}$ is a five-indices label (an explicit construction is given in \cite{Regan:2010cn}). It can then be shown that the estimator \refeq{Estimator} boils down to
\begin{eqnarray}\label{EstimatorPrim2}
\mathcal{E}&=&\frac{1}{N_T}\sum_{m}\alpha_{m}^{\mathcal{Q}}\beta_{m}^{\mathcal{Q}}
\end{eqnarray}
where
\begin{eqnarray}
\beta_{m}^{\mathcal{Q}}=12\int d\hat{n}_1 d\hat{n}_2\int dr_1 dr_2 r_1^2 r_2^2 \mathcal{M}_{m}^{\mathcal{Q}}(\hat{n}_1,\hat{n}_2,r_1,r_2)\,,
\end{eqnarray}
\begin{eqnarray}
\mathcal{M}_{m}^{\mathcal{Q}}&=& N_v(\hat{n}_1,\hat{n}_2,r_1,r_2)  
\Bigg(  M_{p}(\hat{n}_1,r_1)M_{r}(\hat{n}_1,r_1)M_{s}(\hat{n}_2,r_2)M_{u }(\hat{n}_2,r_2) - M^{\rm{uc}}_{p r}(\hat{n}_1,\hat{n}_1,r_1,r_1)M^{\rm{uc}}_{s u }(\hat{n}_2,\hat{n}_2,r_2,r_2) 
\nonumber \\
&-& M^{\rm{uc}}_{p s}(\hat{n}_1,\hat{n}_2,r_1,r_2)M^{\rm{uc}}_{r u }(\hat{n}_1,\hat{n}_2,r_1,r_2) - M^{\rm{uc}}_{p u }(\hat{n}_1,\hat{n}_2,r_1,r_2)M^{\rm{uc}}_{r s}(\hat{n}_1,\hat{n}_2,r_1,r_2)
\Bigg)
\end{eqnarray}
with
\begin{eqnarray}
M^{\rm{uc}}_{p s}(\hat{n}_1,\hat{n}_2,r_1,r_2)&=&\sum_{l_1 m_1}
\frac{Y_{l_1 m_1}(\hat{n}_1)Y_{l_1 m_1}^*(\hat{n}_2)q_p^{l_1}(r_1) q_s^{l_1}(r_2)}{C_{l_1}} ,\nonumber\\
M_{p}(\hat{n}_1,r_1)&=&\sum_{l_1 m_1}\frac{Y_{l_1 m_1}(\hat{n}_1) a_{l_1 m_1}q_p^{l_1}(r_1)}{C_{l_1}},\nonumber\\
N_v(\hat{n}_1,\hat{n}_2,r_1,r_2)&=&\sum_{L M}Y_{L M}^* (\hat{n}_1) Y_{LM}(\hat{n}_2) r_v^L (r_1,r_2)
\end{eqnarray}
and
\begin{eqnarray}
q_p^{l}(r)&=&\frac{2}{\pi}\int dk q_p(k) \Delta_{l}(k)j_{l}(k r),\nonumber\\
r_v^{L}(r_1,r_2)&=&\frac{2}{\pi}\int dK K r_v(K) j_L(K r_1) j_L(K r_2).
\end{eqnarray}
The estimator \refeq{Estimator} has thus been reduced entirely to tractable integrals and sums which can be performed relatively quickly.

\end{document}